\documentclass[12pt]{article}
\usepackage{amsmath,amssymb}
\usepackage{graphicx}
\usepackage{amsmath}
\usepackage{hyperref}
\usepackage{float}
\newcommand {\be}{\begin{equation}}
 \newcommand {\ee}{\end{equation}}
 \newcommand {\bea}{\begin{array}}
 
 \newcommand {\eea}{\end{array}}

\textwidth=6.5in \hoffset=-.75in \voffset=-1in \numberwithin{equation}{section}
\numberwithin{figure}{section}

\def\0{{(0)}}
\def\1{{(1)}}
\def\2{{(2)}}

\def\<{\langle }
\def\>{\rangle }
\def\[{\left[}
\def\]{\right]}

\begin{document}
\begin{titlepage}

\vskip1cm
\begin{center}
{~\\[140pt]{ \LARGE {\textsc{Modular Hamiltonian in flat holography in the framework of  generalized minimal massive gravity   }}}\\[-20pt]}
\vskip2cm

\end{center}
\begin{center}
{M. R. Setare \footnote{E-mail: rezakord@ipm.ir}\hspace{1mm} ,
M. Koohgard \footnote{E-mail: m.koohgard@modares.ac.ir}\hspace{1.5mm} \\
{\small {\em  {Department of Science,\\
 Campus of Bijar, University of Kurdistan, Bijar, Iran }}}}\\
\end{center}
\begin{abstract}

Recently a general prescription for determining the
vacuum modular
flow generator and the corresponding modular Hamiltonian in the BMS-invariant field theories (BMSFTs) have provided by Apolo et. al \cite{Apolo}. According to this paper Einstein gravity in asymptotically flat three-dimensional spacetimes is dual to a BMSFT. In the present paper we extend this study to the generalized minimal massive gravity model (GMMG). So we study some aspects of $Flat_3/BMSFT$ holography in the framework of this model. We derive the modular Hamiltonian for single intervals in thermal and non-thermal BMSFTs and show its holographic dual is the gravitational charge associated with isometries that preserve the asymptotic form of the metric. Also using the modular Hamiltonian we derive the first law of entanglement entropy. Our results for GMMG model reduces to the result for Einstein gravity in $2+1$ dimension, when $\mu , m^2 \to \infty $, and $\sigma=-1$.
These limiting cases, are a place in parameter space of model, where GMMG model reduces to the Einstein gravity.

\end{abstract}
\vspace{1cm}

Keywords: Non-AdS holography, GMMG, Modular Hamiltonian, Modular flow generator, Generalized Rindler prescription

\end{titlepage}

\section{Introduction}	
Holography \cite{tHooft,Sussk} has an important role in the study of entanglement entropy \cite{Calab,Holzhey,Vidal} of quantum field theories.  An evidence for \emph{gauge/gravity duality} is the equality of the modular Hamiltonian \cite{Haag-bk,Casini-1} in the field theory side and the asymptotic isometries of the gravity in the bulk. The modular Hamiltonian is defined through the density matrix of the theory. This Hamiltonian is charge of symmetry group of the system that transforms any operator to itself \cite{Basu}.

The modular Hamiltonian has been used to derive the holographic entanglement entropy in the AdS/CFT correspondence \cite{Casini-2}. Another importance of the modular Hamiltonian is in the derivation of the first law of entanglement \cite{Blanco}. Given importance of this Hamiltonian in AdS/CFT correspondence, we would like to study its role in non-AdS holography. So far the vacuum modular flow generator and the corresponding modular Hamiltonian have been determined in a specific class of non-AdS holographies. In the $Flat_3/BMSFT$ holography as a case of holography beyond AdS/CFT, it is considered that gravity in asymptotically flat three-dimensional spacetimes is dual to a BMS-invariant field theory (BMSFT) \cite{Bagchi-1,Bagchi-5}. In this class of the holography, the flow generated by modular Hamiltonian differs from the modular flow in AdS/CFT \cite{Jiang,Apolo}.

In this paper we consider flat holography in the framework of generalized minimal massive gravity model (GMMG) model. The generalized minimal massive gravity model \cite{Setare 2014} that is realized by adding the CS deformation term, the higher derivative
deformation term, and an extra term to pure Einstein gravity with a negative cosmological constant, is considered in the bulk side of the holography.
 Usually the theories including the terms given by the square of the curvatures have the massive spin
2 mode and the massive scalar mode in addition to the massless graviton. Also the theory has ghosts
due to negative energy excitations of the massive tensor. In \cite{Setare 2014} it is discussed that the GMMG is free
of negative-energy bulk modes, and also avoids the aforementioned
\emph{bulk-boundary unitarity clash}. By a Hamiltonian
analysis one can show that the GMMG model
has no Boulware-Deser ghosts and this model propagate
only two physical modes.
 The group of asymptotic symmetries of asymptotically flat spacetimes at the
future null infinity is the BMS group \cite{Bondi-1,Sach-1,Sach-2}. This symmetry is in correspondence with the symmetry group of the BMSFTs.
An evidence for this holography is equality of the gravitational charge that is related to the asymptotic isometries of the asymptotically flat three dimensional spacetime in GMMG in bulk and the modular Hamiltonian on the field theory side. To prove this correspondence, we find the conserved charge associated with asymptotic isometries of asymptotically flat solutions of  GMMG model. To this end, we consider the metric of asymptotically flat three dimensional spacetime in Bondi gauge. We focus on three special cases of this metric. To compute the modular Hamiltonian we consider BMSFTs in thermal and non-thermal states. Using the generalized Rindler transformation, we compute the modular flow generator \cite{Casini-2,Wen,Jiang,Apolo,Castro} (another method for producing the local modular flow without using the explicit form of the Rindler transformation is proposed in \cite{Wen2}). In this general prescription of the Rindler transformations, some conditions are made on the symmetry transformations $x\to \tilde{x}$ of the coordinates in the field theory. This transformation is invariant under a thermal identification $\tilde{x}^i\sim \tilde{x}^i+i\tilde{\beta}^i$. The vacuum is invariant along this translation and the vacuum state is mapped to a thermal state in the Rindler spacetime.

This paper is organized as follows. In section \ref{sec:2} we introduce briefly the GMMG model. Using the Bondi fall-off conditions, we write the Killing vectors that preserve these conditions. In section \ref{sec:3} we describe the modular Hamiltonian and its relation with the vacuum modular flow generator. Then we find the Killing vectors in two special cases of the asymptotically flat spacetime. In the filed theory side we find the modular Hamiltonian in thermal and non-thermal BMSFTs. We find the first law of holographic entanglement in these cases. In section \ref{sec:4} we provide a summary of the paper and give an overview on this class of $Flat_3/BMSFT$ holography.

\section{Generalized minimal massive gravity in asymptotically flat spaces }\label{sec:2}
Generalized minimal massive gravity (GMMG) is an example of the Chern-Simons-like theories of gravity.
By adding the Chern-Simons term, the higher derivative and an extra term to pure Einstein gravity, the Lagrangian \emph{3-form} of GMMG is given by \cite{Setare 2014}
\begin{eqnarray}\label{L-GMMG}
  L_{GMMG} &=& -\sigma e_aR^a+\frac{\Lambda_0}{6}\epsilon^{abc}e_ae_be_c+h_aT^a+\frac{1}{2\mu}[\omega_ad\omega^a+\frac{1}{3}
  \epsilon^{abc}\omega_a\omega_b\omega_c]\nonumber \\
   &-& \frac{1}{\mathrm{m}^2}[f_aR^a+\frac{1}{2}\epsilon^{abc}e_af_bf_c]+\frac{\alpha}{2}\epsilon^{abc}e_ah_bh_c
\end{eqnarray}

There are four fields that form four flavours of \emph{one-form}, $a^r=\{e,~\omega,~h,~f\}$.  $e$ is dreibein, $h$ and $f$ are the auxiliary \emph{one-form} field and $\omega$ is dualised spin-connection. $T( \omega)$ is Lorentz covariant torsion and $R( \omega)$ is curvature \emph{2-form}. $\mathrm{m}$ is mass parameter of the new massive gravity (NMG) term and $\sigma$ is a sign. The last term of the first line of (\ref{L-GMMG})is the \emph{Lorentz Chern-Simons} term that $\mu$ is mass parameter of that term. $\Lambda_0$ that has dimension mass squared is a cosmological parameter and $\alpha $ is a dimensionless parameter. The last term in the Lagrangian of GMMG (\ref{L-GMMG}) is an extra  term that is added to the model.

The following fall-off conditions is considered for the asymptotically $2+1$-dimensional flat spacetime \cite{Setare-Adami Nucl 2018,Detournay 2017}
\begin{eqnarray}\label{B.C.}
  g_{uu} &=& \mathcal{M}(\phi)+\mathcal{O}(r^{-2}) \nonumber\\
  g_{ur} &=& -e^{\mathcal{A}(\phi)}+\mathcal{O}(r^{-2}) \nonumber\\
  g_{u\phi} &=& \mathcal{N}(u,\phi)+\mathcal{O}(r^{-1}) \nonumber\\
  g_{rr} &=& \mathcal{O}(r^{-2}) \nonumber\\
  g_{r\phi} &=& -e^{2\mathcal{A}(\phi)}\mathcal{E}(u,\phi)+\mathcal{O}(r^{-1}) \nonumber\\
  g_{\phi\phi} &=& e^{\mathcal{A}(\phi)}r^2+\mathcal{E}(u,\phi)[2\mathcal{N}(u,\phi)-
  \mathcal{M}(\phi)\mathcal{E}(u,\phi)]+\mathcal{O}(r^{-1})
\end{eqnarray}
where
\begin{eqnarray}\label{N,E}
  \mathcal{N}(u,\phi) &=& \mathcal{L}(\phi)+\frac{u}{2}\partial_{\phi}\mathcal{M}(\phi)\nonumber \\
  \mathcal{E}(u,\phi) &=& \mathcal{B}(\phi)+u\partial_{\phi}\mathcal{A}(\phi)
\end{eqnarray}

In the above equations $\mathcal{A}(\phi)$, $\mathcal{L}(\phi)$, $\mathcal{B}(\phi)$ and $\mathcal{M}(\phi)$ are arbitrary functions and $r$, $u$ and $\phi$ are the coordinates. The conditions admit the Bondi-Sachs  fall-off conditions at future null infinity. The metric transformations generated by vector field $\xi$ are obtained by $\delta_{\xi}g_{\mu\nu}=\mathcal{\L}_{\xi}g_{\mu\nu}$. $\mathcal{\L}_{\xi}$ is the Lie derivative along $\xi$. The Killing vector fields that preserve the boundary conditions invariant are as follows \cite{Setare-Adami Nucl 2018,Detournay 2017}
\begin{eqnarray} \label{Killing GMMG}
   \xi^u & = & \alpha(u,\phi)-\frac{1}{r}e^{-\mathcal{A}(\phi)}\mathcal{E}(u,\phi)\beta(u,\phi)+\mathcal{O}(r^{-2})\nonumber \\
    \xi^r &=& r X(\phi)+e^{-\mathcal{A}(\phi)}[\mathcal{E}(u,\phi)\partial_{\phi}X(\phi)-\partial_{\phi}\beta(u,\phi)]\nonumber \\
     & +& \frac{1}{r}e^{-2\mathcal{A}(\phi)}\beta(u,\phi)[\mathcal{N}(u,\phi)-\mathcal{M}(\phi)\mathcal{E}(u,\phi)]
     +\mathcal{O}(r^{-2}),\nonumber \\
     \xi^{\phi} &=& Y(\phi)+\frac{1}{r}e^{-\mathcal{A}(\phi)}\beta(u,\phi)+\mathcal{O}(r^{-2}),
\end{eqnarray}
where
\begin{eqnarray}\label{alfa-beta}
  \alpha(u,\phi) &=& T(\phi)+u\partial_{\phi}Y(\phi), \nonumber \\
  \beta(u,\phi) &=& Z(\phi)+u\partial_{\phi}X(\phi), \nonumber \\
   \beta(u,\phi) &=& - \partial_{\phi}\alpha(u,\phi)
\end{eqnarray}

$T(\phi)$, $Y(\phi)$, $Z(\phi)$ and $X(\phi)$ are arbitrary functions of $\phi$ that are related by (\ref{alfa-beta}). The conserved charges of asymptotically flat spacetime corresponds to the Killing vector field (\ref{Killing GMMG}) are obtained in \cite{Setare-Adami Nucl 2018}. Using the Fourier modes of the charges, it has been shown the asymptotic symmetry algebra is a semidirect product of a $BMS_3$ algebra and two $U(1)$ current algebras \cite{Setare-Adami Nucl 2018}. These charges are not only for the Killing vector fields are satisfied by spacetime everywhere but also these are admitted for the Killing vectors in asymptotic configuration.
The authors in \cite{Setare-Adami Nucl 2018} have used the quasi-local conserved charge perturbation associated with a field dependent vector field $\xi$ in GMMG model for asymptotically flat spacetimes and they have used an integration over the one-parameter path on the solution space from the charge perturbation to find the following conserved charges
\begin{equation}\label{Total charge}
  Q(T,X,Y,Z)=M(T)+J(X)+L(Y)+P(Z),
\end{equation}
where

\begin{eqnarray}
  M(T) &=& -\frac{1}{16\pi G}(\sigma+\frac{\alpha H}{\mu}+\frac{F}{\mathrm{m}^2})\int_{0}^{2\pi}T(\phi)\mathcal{M}(\phi)
  d\phi, \label{M- charge}\\
  J(X) &=& \frac{1}{8\pi G}\int_{0}^{2\pi}X(\phi)\big[(\sigma+\frac{\alpha H}{\mu}+\frac{F}{\mathrm{m}^2})
  \partial_{\phi}\mathcal{B}(\phi)-\frac{1}{2\mu}\partial_{\phi}\mathcal{A}(\phi)\big] d\phi,\label{J-charge} \\
  L(Y) &=& -\frac{1}{8\pi G} \int_{0}^{2\pi}Y(\phi)\bigg\{(\sigma+\frac{\alpha H}{\mu}+\frac{F}{\mathrm{m}^2})\mathcal{L}(\phi)\nonumber    \\
   &-& \frac{1}{4\mu}\big[2\mathcal{M}(\phi)+(\partial_{\phi}\mathcal{A}(\phi))^2-2\partial_{\phi}^2\mathcal{A}(\phi) \big] \bigg\}d\phi, \label{L-charge} \\
   P(Z) &=& \frac{1}{8\pi G}(\sigma+\frac{\alpha H}{\mu}+\frac{F}{\mathrm{m}^2})\int_{0}^{2\pi}Z(\phi)\partial_{\phi}\mathcal{A}(\phi)d\phi. \label{P-charge}
\end{eqnarray}

$H$ and $F$ are constant parameters.

\section{Flat holography in the framework of GMMG }\label{sec:3}
The BMS group is the symmetry group of the asymptotically flat spacetime in three dimensions \cite{Bondi-1,Sach-1,Sach-2,Barni-Comp}. This group is an
infinit-dimensional one in $3$ and $4$-dimensions . The BMS algebra in three dimensions is the same as the symmetry algebra of a BMS-invariant field theory (BMSFT) in two dimensions. This motivate us to a $Flat_3/BMSFT$ duality \cite{Bagchi-1,Barni-Troes}.
An evidence for this correspondence is computation of the entanglement entropy \cite{Bagchi-2,Basu}. The entanglement entropy of 2d BMSFT at the boundary is the same as the entropy in the gravity side. Another evidence for the correspondence is reproducing the Bekenstein-Hawking entropy of Flat space cosmologies using a modified Cardy formula for the BMSFTs \cite{Barni-1,Bagchi-3}. Anyone who is interested could see in \cite{Bagchi-4} to a review a number of applications of this holography.

In our previous work \cite{we}, we considered the GMMG model in the asymptotically flat space in three cases: a flat space cosmological geometry (FSC), a null orbifold and a Poincare patch. The 2d field theory on the boundary was considered as a BMSFT. Using the Cardy formula, the entaglement entropy of the field theory side was computed and was compared with the Beckenestein-Hawking entropy of the garvity side. The entropy in two sides of the GMMG/BMSFT correspondence is another evidence for $Flat_3/BMSFT$ holography conjecture. In this section we study the modular Hamiltonian in both sides of the correspondence and derive the first law of the entanglement entropy.

The modular Hamiltonian has an important role in holography and black hole physics. On a subregion $\mathcal{A}$ of theory in the boundary, the density matrix $\rho_{\mathcal{A}}$ is described through its modular Hamiltonian as follows \cite{Haag-bk,Casini-1}
\begin{equation}\label{H mod : int}
  \mathcal{H}_{mod}=-\log\rho_{\mathcal{A}}
\end{equation}
where the modular Hamiltonian is generically independent of some local expressions constructed with the fields on $\mathcal{A}$. This Hamiltonian generically is not a local operator \cite{Casini-2}. The modular Hamiltonian is charge of a symmetry of the system through the unitary operator $U(s)=e^{-i\mathcal{H}_{mod}s}$. This symmetry group transforms any operator in the causal domain of $\mathcal{A}$ into itself as follows \cite{Haag-bk,Casini-2,Boer}
\begin{equation}\label{H mod:cp}
  e^{i\mathcal{H}_{mod}s}\mathcal{O}(\mathcal{A})e^{-i\mathcal{H}_{mod}s}=\mathcal{O}(\mathcal{A})
\end{equation}
where $\mathcal{O}(\mathcal{A})$ is the algebra of operators in $\mathcal{A}$. This internal time transformations on the charge algebra in $\mathcal{A}$ is a one-parameter group. This group is called the modular flow corresponding to $\mathcal{A}$. Based on the generalized Rindler method \cite{Wen,Jiang}, the modular flow generator can be defined as a linear combination of the vacuum symmetry generators $h_i$ as follows
\begin{equation}\label{flow gen:int}
  \zeta=\Sigma_i a_ih_i
\end{equation}
where $a_i$ are some arbitrary constants and the relation (\ref{flow gen:int}) is just true for the local modular flows. The vacuum symmetry generators $h_i$ preserve the vacuum of the field theory invariant. Corresponding to the dictionary of holography, the isometry group of the gravity theory in the bulk $H$ is dual to the isometry group of the vacuum state of the field theory on the boundary. There is the following correspondence between the vacuum symmetry generators $h_i$ and the Killing vectors $H_i$ of diffeomorphism in the bulk
\begin{equation}\label{H=h}
  H_i|_{\partial M}=h_i
\end{equation}
where $H_i|_{\partial M}$ are defined on a surface $\partial M$. According to (\ref{H=h}), we find the following correspondence between the modular flow generator in the bulk and the modular flow generator on the boundary
\begin{equation}\label{xi=zeta}
  \xi|_{\partial M}=\zeta
\end{equation}
where
\begin{equation}\label{xi:gen}
\xi=\Sigma_ia_iH_i.
\end{equation}

The coefficients $a_i$ in (\ref{xi:gen}) are the same as the parameters in (\ref{flow gen:int}). In the Bondi gauge, we can re-parameterize the fall-off conditions (\ref{B.C.}) as follows \cite{Barni-Troes}
\begin{equation}\label{flat space-2}
  ds^2=Mdu^2-2dudr+Jdud\phi+r^2d\phi^2
\end{equation}
where $M$ and $J$ can be regarded as the mass and the angular momentum of the spacetime respectively. Comparing Eqs. (\ref{flat space-2}) and (\ref{B.C.}), we have the following identification
\begin{eqnarray} \label{Ids. 1}
  \mathcal{M}(\phi) &=& M \nonumber\\
  \mathcal{A} &=& 0 \nonumber\\
  2\mathcal{N}(u,\phi) &=& J\nonumber \\
  \mathcal{B} &=& 0\nonumber \\
  \mathcal{E}(u,\phi) &=& 0
\end{eqnarray}

We are especially interested in three cases corresponding to the parameters $M$ and $J$ values:

(i) $M=-1$ and $J=0$: In this case the background spacetime is converted to the Minkowski spacetime in three dimensions (called global Minkowski slotuion). This metric has the following form
\begin{equation}\label{Mink}
  ds^2=-du^2-2dudr+r^2d\phi^2
\end{equation}

(ii) $M=J=0$: The metric has the following form
\begin{equation}\label{Poin}
  ds^2=-2dudr+r^2d\phi^2,
\end{equation}
that is the Poincare patch in the $AdS_3$ spacetime. The manifold is a null orbifold that is defined by a null boost \cite{Horowitz}.

(iii)$M>0$: In this case we have a flat space cosmological geometry (FSC) that is
time-dependent solution of Einstein equation in three dimensions of spacetime \cite{we,FSC-1,FSC-2}.

Using the identifications (\ref{Ids. 1}), the Killing vector fields (\ref{Killing GMMG}) and (\ref{alfa-beta}) that preserve the Bondi boundary conditions invariant are as follows
\begin{eqnarray} \label{Killing GMMG-2}
  \xi^u &=& T(\phi)+u\partial_{\phi}Y(\phi)\nonumber \\
  \xi^r &=& -\frac{J}{2r}\partial_{\phi}T(\phi)-\frac{J}{2r}u\partial_{\phi}^2Y(\phi)+u\partial_{\phi}^3Y(\phi)+\partial_{\phi}^2T(\phi)
  -r\partial_{\phi}Y(\phi)\nonumber \\
  \xi^{\phi} &=& Y(\phi)- \frac{1}{r} \partial_{\phi}T(\phi)-\frac{u}{r}\partial_{\phi}^2Y(\phi),
\end{eqnarray}

 By substituting $\mathcal{A}=0$ and $\mathcal{B}=0$ into the conserved charges GMMG model (\ref{M- charge}), (\ref{J-charge}), (\ref{L-charge}) and (\ref{P-charge}) for asymptotically flat spacetime
 we have the following form for the charges
 \begin{eqnarray}
  M(T) &=& -\frac{1}{16\pi G}(\sigma+\frac{\alpha H}{\mu}+\frac{F}{\mathrm{m}^2})\int_{0}^{2\pi}T(\phi)\mathcal{M}(\phi)
  d\phi, \label{M- charge2}\\
  J(X) &=& 0,\label{J-charge2} \\
  L(Y) &=& -\frac{1}{8\pi G} \int_{0}^{2\pi}Y(\phi)\bigg\{(\sigma+\frac{\alpha H}{\mu}+\frac{F}{\mathrm{m}^2})\mathcal{L}(\phi)-\frac{1}{2\mu}\mathcal{M}(\phi)\bigg\}d\phi \label{L-charge2} \\
   P(Z) &=& 0 \label{P-charge2}
\end{eqnarray}
By adding these results together we obtain following expression
\begin{eqnarray} \label{F-charge GMMG}
  Q &=& M(T)+J(X)+L(Y)+P(Z)\nonumber \\
   &=&  -\frac{1}{16\pi G}(\sigma+\frac{\alpha H}{\mu}+\frac{F}{\mathrm{m}^2})\int_{0}^{2\pi}\bigg\{T(\phi)\mathcal{M}(\phi)+2Y(\phi)\mathcal{L}(\phi)\bigg\}d\phi\nonumber \\
   &+& \frac{1}{16\pi G \mu}\int_{0}^{2\pi}Y(\phi)\mathcal{M}(\phi)d\phi
\end{eqnarray}

This is the gravitational charge corresponds to the modular flow generator (\ref{Killing GMMG-2}). The infinitesimal variation corresponding to this charge is as follows
\begin{eqnarray} \label{delta F-charge GMMG}
  \delta Q &=& \delta M(T)+\delta J(X)+\delta L(Y)+\delta P(Z)\nonumber \\
   &=&  -\frac{1}{16\pi G}(\sigma+\frac{\alpha H}{\mu}+\frac{F}{\mathrm{m}^2})\int_{0}^{2\pi}\bigg\{T(\phi)\delta \mathcal{M}(\phi)+2Y(\phi)\delta \mathcal{L}(\phi)\bigg\}d\phi\nonumber \\
   &+& \frac{1}{16\pi G \mu}\int_{0}^{2\pi}Y(\phi)\delta \mathcal{M}(\phi)d\phi
\end{eqnarray}

It is interesting and as one expect, the above result reduces to the expression (3.51)  in the paper \cite{Apolo} for Einstein gravity in $2+1$ dimension, when $\mu , \mathrm{m}^2 \to \infty $ and $\sigma=-1$. In these limiting cases, GMMG model reduces to the Einstein gravity. We use this result to show an evidence for the $flat_3/BMSFT$ correspondence. In the field theory side, the story is a little different. To compute the vacuum modular Hamiltonian for a single interval in BMSFTs on the plane, we first find the modular flow generator from the generalized Rindler transformation \cite{Casini-2,Jiang,Apolo,Castro}. In this prescription a symmetry transformation of the vacuum state is used to map an interval $\mathcal{I}$ (and its causal domain $\mathcal{D}$) to a Rindler spacetime. If \textbf{G} is a symmetry group that preserves the field theory invariant, a subgroup $\textbf{H}\subset\textbf{G}$ is a symmetry group that preserves the vacuum state invariant. In the Rindler spacetime, the generator of time-like translations is identified with the modular Hamiltonian \cite{Castro,Apolo}.

In a generalized Rindler prescription \cite{Castro,Apolo}, it could be possible to find the modular flow generator without an explicit expression for the Rindler transformations. For this end, the modular flow generator is considered as a linear combination of the vacuum symmetry generators similar to \ref{flow gen:int}. There is a condition to determine the coefficients $a_i$ and that is the invariance of the causal domain $\partial\mathcal{D}$ under the modular flow generator. To determine the overall normalization of the modular flow generator, it should be assumed that $e^{i \zeta}$ maps any point back to itself.

To apply the generalized Rindler method, we consider an interval $\mathcal{I}$ on the vacuum state as follows
\begin{equation}\label{interval}
  \partial\mathcal{I}=\{(u_-\phi_-),(u_+,\phi_+)\}
\end{equation}
where $u$ and $\phi$ are the coordinates on the plane and the interval $\mathcal{I}$ is identified with its endpoints. The modular flow generator as linear combination of vacuum symmetry generators $l_j$ and $m_j$ can be write as follows
\begin{equation}\label{zeta-1}
  \zeta=\zeta^u\partial_u+\zeta^\phi\partial_\phi=\Sigma_{j=-1}^1(a_jl_j+b_jm_j),
\end{equation}
where $l_j$ and $m_j$ are defined as follows
\begin{eqnarray}
  l_j &=& -\phi^{j+1}\partial_\phi-(j+1)\phi^ju\partial_u,\nonumber \\
  m_j &=& -\phi^{j+1}\partial_u \label{l&m1}
\end{eqnarray}
and satisfy the BMS algebra
\begin{eqnarray}
  [l_j,l_k] &=& (j-k)l_{j+k}, \\ \label{l&l}
  [l_j,m_k] &=& (j-k)m_{j+k}, \\ \label{l&m2}
  [m_j,m_k] &=& 0 \label{m&m}
\end{eqnarray}

To determine the coefficients $a_j$ and $b_j$, it is required to assume the boundary of the causal domain $\partial\mathcal{D}$ is invariant under the modular flow. By this requirement, the modular flow generator vanishes at the endpoints of the interval $\mathcal{I}$ (\ref{interval}) as follows
\begin{eqnarray} \label{zeta right}
  \zeta_{(u_+,\phi_+)} &=& 0 \nonumber \\
  &=& a_+l_++a_0l_0+a_-l_-+b_+m_++m_0b_0+b_-m_-
\end{eqnarray}
 \begin{eqnarray} \label{zeta left}
  \zeta_{(u_-,\phi_-)} &=& 0 \nonumber \\
  &=& a_+l_++a_0l_0+a_-l_-+b_+m_++m_0b_0+b_-m_-
\end{eqnarray}

Substituting $l_j$ and $m_j$ definitions (\ref{l&m1}) into the above equations, we find four relations for $a_j$ and $b_j$ as follows
\begin{eqnarray}\label{coef}
  a_+\phi_+^2+a_0\phi_++a_- &=& 0,  \\
  a_+\phi_-^2+a_0\phi_-+a_- &=& 0, \label{b} \\
  2a_+\phi_+u_++a_0u_++b_+\phi_+^2+b_0\phi_++b_- &=& 0, \label{c} \\
  2a_+\phi_-u_-+a_0u_-+b_+\phi_-^2+b_0\phi_-+b_- &=& 0. \label{d}
\end{eqnarray}
These relations are corresponding to the coefficients of $\partial_u$ and $\partial_\phi$ when the $l_j$ and $m_j$ definitions (\ref{l&m1}) is substituted into the equations (\ref{zeta right}) and (\ref{zeta left}). Using Eqs. (\ref{coef}) and (\ref{b}), we have the following expressions for the coefficients $a_j$
\begin{equation}\label{aj}
  (a_+,a_0,a_-)=a_+\big(1,-(\phi_++\phi_-),\phi_+\phi_-\big).
\end{equation}
The solution of the Eqs. (\ref{c}) and (\ref{d}) for the coefficients $b_j$ is as follows
\begin{equation}\label{bj}
  (b_+,b_0,b_-)=b_+\big(1,-(\phi_++\phi_-),\phi_+\phi_-\big)+a_+\big(0,-(u_-+u_+),u_+\phi_-+u_-\phi_+\big)
\end{equation}
To find $a_+$ and $a_-$, the following differential equations are considered \cite{Apolo}
\begin{equation}\label{traject}
  \partial_su(s)=\zeta^u,~~~~\partial_s\phi(s)=\zeta^\phi,
\end{equation}
where $(\phi(s),u(s))$ is the trajectory under the modular flow $e^{s\zeta}$. By solving the differential equations (\ref{traject}) and applying the conditions $\phi(s)=\phi(s+i)$ and $u(s)=u(s+i)$, $a_+$ and $a_-$ are found as follows  \cite{Apolo}
\begin{eqnarray}\label{ab+}
  a_+ &=& \frac{2\pi}{\phi_+-\phi_-},\nonumber \\
  b_+ &=& -\frac{2\pi(u_+-u_-)}{(\phi_+-\phi_-)^2}.
\end{eqnarray}

Substituting the results for $a_+$ and $a_-$ into eqs. (\ref{aj}) and (\ref{bj}), the coefficients are computed as follows
\begin{eqnarray}\label{aj2}
  (a_+,a_0,a_-) &=& \frac{2\pi}{\phi_+-\phi_-}\big(1,-(\phi_++\phi_-),\phi_+\phi_-\big) \\
  (b_+,b_0,b_-) &=& \frac{2\pi}{(\phi_+-\phi_-)^2}\big(u_--u_+,2u_+\phi_--2u_-\phi_+,u_-\phi_+^2-u_+\phi_-^2\big)\label{bj2}
\end{eqnarray}

By the coefficients (\ref{aj2}) and (\ref{bj2}), we have the modular flow generator $\zeta$. We now can compute the vacuum modular Hamiltonian for an interval $\mathcal{I}$ in BMSFT on the plane. To find this Hamiltonian, the vacuum symmetry generators $l_j$ and $m_j$ should be replaced with the corresponding conserved charges $\mathcal{L}_j$ and $\mathcal{M}_j$. These charges are defined as follows

\begin{eqnarray}
  \mathcal{L}_j &=& \frac{1}{2\pi}\int d\phi \phi^{j+1}\mathcal{J}(\phi),\nonumber \\
  \mathcal{M}_j &=& \frac{1}{2\pi}\int d\phi \phi^{j+1}\mathcal{P}(\phi) \label{chs}
\end{eqnarray}
where $\mathcal{J}(\phi)$ and $\mathcal{P}(\phi)$ are the conserved currents associated with $\mathcal{L}_j$ and $\mathcal{M}_j$, respectively.

Replacing the symmetry generators $l_j$ and $m_j$ with the charges (\ref{chs}) in (\ref{zeta-1}), we find the modular Hamiltonian as follows
\begin{eqnarray}\label{H in FT}
  \mathcal{H}_{\zeta} &=& \Sigma_{j=-1}^1(a_j\mathcal{L}_j+b_j\mathcal{M}_j)\nonumber \\
   &=& -\frac{1}{2\pi}\int_{\phi_-}^{\phi_+}(T(\phi)\mathcal{P}(\phi)+Y(\phi)\mathcal{J}(\phi))
\end{eqnarray}
where by substituting the coefficients (\ref{aj2}) and (\ref{bj2}) into the first line of the above equation, $T(\phi)$ and $Y(\phi)$ are found as follows
\begin{eqnarray}\label{T}
  T(\phi) &=& \frac{2\pi[u_+(\phi-\phi_-)^2-u_-(\phi-\phi_+)^2]}{(\phi_+-\phi_-)^2} \\
  Y(\phi) &=& -\frac{2\pi(\phi-\phi_-)(\phi-\phi_+)}{\phi_+-\phi_-} \label{Y}
\end{eqnarray}
where the functions $T(\phi)$ and $Y(\phi)$ are non-vanishing only within the causal domain $\mathcal{D}$. So far we have calculated the modular Hamiltonian on the BMSFT on a plane in (\ref{H in FT}) and the gravitational charge corresponding to the modular flow generator in (\ref{F-charge GMMG}). A consistency check on these results is that these results satisfy the first law of entanglement entropy. We now show that the modular Hamiltonian (\ref{H in FT}) is the same as the gravitational charge (\ref{F-charge GMMG}). To this end, we compare the charge algebras on two sides of the holography. By introducing the following Fourier modes of the conserved charges (\ref{M- charge2}), (\ref{J-charge2}), (\ref{L-charge2}), (\ref{P-charge2}) and (\ref{F-charge GMMG}) of the asymptotically flat spacetime in GMMG
\begin{eqnarray}\label{fourier1}
  M_n &=& Q(e^{in\phi},0,0,0)=M(e^{in\phi}),\nonumber \\
  J_n &=& Q(0,e^{in\phi},0,0)=J(e^{in\phi}),\nonumber \\
  L_n &=& Q(0,0,e^{in\phi},0)=L(e^{in\phi}),\nonumber \\
  P_n &=& Q(0,0,0,e^{in\phi})=P(e^{in\phi}),
\end{eqnarray}
we find the non-zero Fourier modes as follows
\begin{eqnarray}
  M_n(T) &=& -\frac{1}{16\pi G}(\sigma+\frac{\alpha H}{\mu}+\frac{F}{\mathrm{m}^2 })\int_{0}^{2\pi}T(\phi)\mathcal{M}(\phi)e^{in\phi}
  d\phi, \label{Mm- charge2}\\
  L_n(Y) &=& -\frac{1}{8\pi G} \int_{0}^{2\pi}Y(\phi)\bigg\{(\sigma+\frac{\alpha H}{\mu}+\frac{F}{\mathrm{m}^2})\mathcal{L}(\phi)-\frac{1}{2\mu}\mathcal{M}(\phi)\bigg\}e^{in\phi}d\phi \label{Lm-charge2}
\end{eqnarray}
where $m$ indicates the Fourier modes and $\mathrm{m}$ is the mass parameter of NMG.

It could be proved that the gravitational charges (\ref{Mm- charge2}) and (\ref{Lm-charge2}) satisfy three-dimensional BMS algebra as follows \cite{Setare-Adami Nucl 2018}
\begin{eqnarray}\label{BMS algebra}
  \[L_m,L_n\] &=& (m-n)L_{m+n}+\frac{c_L}{12}m^3\delta_{m+n,0}\nonumber \\
  \[M_m,L_n\] &=& (m-n)M_{m+n}+\frac{c_M}{12}m^3\delta_{m+n,0}\nonumber \\
  \[M_m,M_n\] &=& 0
\end{eqnarray}
where the $c_L$ and $c_M$ are the central extensions as follows
\begin{eqnarray}\label{cL,cM}
  c_L &=& \frac{3}{G\mu}\nonumber \\
  c_M &=& -\frac{3}{G}(\sigma+\frac{\alpha H}{\mu}+\frac{F}{\mathrm{m}^2})
\end{eqnarray}
The Fourier modes of the $\mathcal{J}(\phi)$ and $\mathcal{P}(\phi)$ currents (\ref{chs}) are as follows
\begin{eqnarray}\label{fourier2}
  \mathcal{L}_m &=& -\frac{1}{2\pi}\int d\phi e^{im\phi}\mathcal{J}(\phi),\nonumber \\
  \mathcal{M}_m &=& -\frac{1}{2\pi}\int d\phi e^{im\phi}\mathcal{P}(\phi)
\end{eqnarray}
These charges satisfy the same BMS algebra of the asymptotically flat spacetime in GMMG (\ref{BMS algebra}) where
\begin{equation}\label{charge dict1}
  L_m\to \mathcal{L}_m, ~~~~~ M_m\to \mathcal{M}_m
\end{equation}

This correspondence between the bulk and the boundary symmetry algebras is an indication of $flat_3/BMSFT$ holography. By this correspondence, we find the following dictionary between the bulk and the boundary variables
\begin{eqnarray}\label{var.rel}
  \frac{\mathcal{P}}{2\pi} &=& \frac{1}{16\pi G}(\sigma+\frac{\alpha H}{\mu}+\frac{F}{\mathrm{m}^2})\mathcal{M}(\phi)\nonumber \\
  \frac{\mathcal{J}}{2\pi} &=& \frac{1}{8\pi G}\bigg\{(\sigma+\frac{\alpha H}{\mu}+\frac{F}{\mathrm{m}^2})\mathcal{L}(\phi)-\frac{1}{2\mu}\mathcal{M}\bigg\}
\end{eqnarray}

Using this holographic dictionary, the gravitational charge (\ref{F-charge GMMG}) is the same as the modular Hamiltonian (\ref{H in FT}) on the field theory side as follows
\begin{equation}\label{match1}
  \delta Q\[g\]=\delta \langle\mathcal{H}_{\zeta}\rangle=\delta \langle\mathcal{H}_{mod}\rangle
\end{equation}

In \cite{Apolo-2}, the authors have proposed a general prescription for matching the holographic entanglement entropy and the gravitational charge evaluated on an asymptotically flat space. So we have the following result for the first law of entanglement entropy in $flat_3/BMSFT$
\begin{equation}\label{match1}
  \delta S_{EE}=\delta Q\[g\]=\delta \langle\mathcal{H}_{\zeta}\rangle=\delta \langle\mathcal{H}_{mod}\rangle
\end{equation}

To find this result, we have considered the Poincare patch (\ref{Poin}) in the gravity side of the holography. In the field theory side, we have had a BMSFT with zero temperature on the plane (\ref{interval}). We could generalize the result to more general cases in the bulk and boundary of the $flat_3/BMSFT$ correspondence. In the boundary, we consider the thermal BMSFT that is defined on a thermal cylinder that has the following thermal identification of the coordinates
\begin{equation}\label{therm. cyl}
  (u,\phi)\sim (u,,\phi+2\pi)\sim (u+i\beta_u,\phi-i\beta_{\phi})
\end{equation}
where
\begin{equation}\label{beta1}
  \beta_u=\frac{\pi J}{M^{3/2}},~~~~ \beta_{\phi}=\frac{2\pi}{\sqrt{M}}.
\end{equation}

Now, we consider the Flat Space Cosmological solution (FSC) that is third case of general asymptotically flat space time solution of GMMG, this case is considered with the following metric,
\begin{equation}\label{FSC}
  ds^2=Mdu^2-2dudr+Jdud\phi+r^2d\phi^2, ~~~~ M>0.
\end{equation}

To find the modular flow generator, the interval $\mathcal{I}$ at the boundary is parameterized
 in terms of the endpoints as follows
 \begin{equation}\label{therm. cyl.2}
   \partial\mathcal{I}=\{(u_-,\phi_-),(u_+,\phi_+)\},~~~~~ l_u\equiv u_+-u_-,~~l_{\phi}\equiv\phi_+-\phi_-.
 \end{equation}

The generalized Rindler method could be used to find the modular flow generator. The modular flow generator can be written as follows
\begin{eqnarray}\label{zeta-therm}
  \zeta &=& \zeta^u\partial_u+\zeta^\phi\partial_\phi=\Sigma_{j=-1}^1a_jl_j+b_jm_j\nonumber\\
  &=& [T(\phi)+uY'(\phi)]\partial_u+Y(\phi)\partial_\phi
\end{eqnarray}

To write the $2nd$ line of the above equation, we use the following Killing vectors of FSC in $r\to\infty$ limit \cite{Jiang}
\begin{eqnarray}\label{Kil-FSC-2}
  l_1 &=& -e^{-\sqrt{M}\phi}\bigg(\frac{(\sqrt{M}r_c\phi+Mu+r_c)}{M},
  -r,-\frac{1}{\sqrt{M}}\bigg),\nonumber \\
  l_0 &=& (-\frac{r_c}{M},0,\frac{1}{\sqrt{M}}),\nonumber \\
  l_{-1} &=& e^{\sqrt{M}\phi}\bigg(\frac{(\sqrt{M}r_c\phi+Mu-r_c)}{M},
  -r,\frac{1}{\sqrt{M}}\bigg),\nonumber  \\
  m_1 &=& -e^{-\sqrt{M}\phi}\big(\frac{1}{\sqrt{M}},\sqrt{M},0\big),\nonumber \\
  m_0 &=& (-\frac{1}{\sqrt{M}},0,0),\nonumber \\
  m_{-1} &=& e^{\sqrt{M}\phi}\big(-\frac{1}{\sqrt{M}},-\sqrt{M},0\big).
\end{eqnarray}

To find the explicit form of $Y(\phi)$, we expand the first line of (\ref{zeta-therm}) using the Killing vectors (\ref{Kil-FSC-2}), then we obtain
\begin{equation}\label{Y-2}
  Y=b_1(\frac{e^{-\sqrt{M}\phi}}{\sqrt{M}})+b_0(\frac{1}{\sqrt{M}})+b_{-1}(\frac{e^{\sqrt{M}\phi}}{\sqrt{M}})
\end{equation}

Fixing the center point of the interval at $\phi=0$, we set $b_1=b_{-1}$. $b_0$ and $b_1$ could be considered as follows \cite{Jiang}
\begin{eqnarray}
  b_0 &=& \frac{\pi}{\tilde{\beta}_{\phi}}\bigg(\tanh\big(\frac{\pi l_{\phi}}{2\beta_{\phi}}\big)+\coth\big(\frac{\pi l_{\phi}}{2\beta_{\phi}}\big)\bigg)\nonumber \\
  b_1 &=& -\frac{\pi}{\tilde{\beta}_{\phi}}\text{csch}\big(\frac{\pi l_{\phi}}{\beta_{\phi}}\big) \label{bb}
\end{eqnarray}

Substituting the definitions (\ref{bb}) into (\ref{Y-2}) and using some manipulations, we find the following form for $Y(\phi)$
\begin{equation}\label{Y-3}
  Y(\phi)=\frac{2\pi}{\sqrt{M}\sinh\big(\frac{\sqrt{M}l_{\phi}}{2}\big) }
  \big[\cosh\big(\frac{\sqrt{M}l_{\phi}}{2}\big)-\cosh(\sqrt{M}\Delta \phi)\big]
\end{equation}
where we use the following identifications
\begin{equation}\label{ids-2}
  \Delta\phi\equiv \phi-(\phi_++\phi_-)/2.
\end{equation}

A similar calculation could be done to find $T(\phi)$. To this end, we consider $d_1=d_{-1}$ and we have the following definitions \cite{Jiang}
\begin{eqnarray}
  d_1 &=& -\frac{\pi~\text{csch}(\pi l_{\phi}/\beta_{\phi})\big[-\beta^2_{\phi}\tilde{\beta}_u
  +\pi(l_{\phi}\beta_u+l_u\beta_{\phi})\tilde{\beta}_{\phi}
  \coth(\pi l_{\phi}/\beta_{\phi})\big]}{\beta^2_{\phi}\tilde{\beta}^2_{\phi}}, \nonumber \\
  d_0 &=& \frac{\pi~\text{csch}^2(\pi l_{\phi}/\beta_{\phi})\big[-\beta^2_{\phi}\tilde{\beta}_u \sinh(2\pi l_{\phi}/\beta_{\phi})
  +2\pi(l_{\phi}\beta_u+l_u\beta_{\phi})\tilde{\beta}_{\phi}
  \big]}{\beta^2_{\phi}\tilde{\beta}^2_{\phi}} \label{dd}
\end{eqnarray}
We find $T(\phi)$ as follows
\begin{eqnarray}
  T(\phi) &=& \frac{\pi}{2M\sinh\big(\frac{\sqrt{M}l_{}\phi}{2}\big)}\bigg\{(Jl_{\phi}+2l_uM)
  \big[\coth\big(\frac{\sqrt{M}l_{\phi}}{2}\big)\cosh(\sqrt{M}\Delta\phi)-\text{csch}\big(\frac{\sqrt{M}l_{\phi}}{2}\big)
  \big]
  \nonumber \\
   &+& \frac{2J}{\sqrt{M}}\big[\cosh(\sqrt{M}\Delta\phi)-\cosh\big(\frac{\sqrt{M}l_{\phi}}{2}\big)
   \big]-2J\Delta\phi\sinh(\sqrt{M}\Delta\phi)\bigg\}
\end{eqnarray}

A similar computation to that we did to find the modular Hamiltonian (\ref{H in FT}), can be done to find the following result for the modular Hamiltonian for a single interval on a thermal BMSFT
\begin{equation}\label{H in FT-therm}
  \mathcal{H}_{\zeta} = -\frac{1}{2\pi}\int_{\phi_-}^{\phi_+}\big(T(\phi)\mathcal{P}(\phi)+Y(\phi)\mathcal{J}(\phi)\big)
\end{equation}

There is the correspondence (\ref{H=h}) between the vacuum symmetry generators $h_i$ at the boundary and the Killing vectors $H_i$ in the bulk. The Killing vectors in the bulk are given in terms of the $T(\phi)$ and $Y(\phi)$ by (\ref{Killing GMMG-2}). This form of the Killing vectors that we have derived can be used in the FSC geometry. The gravitational charge associated with the Killing vectors is then given by
\begin{eqnarray} \label{F-charge GMMG2}
  Q &=& M(T)+J(X)+L(Y)+P(Z)\nonumber \\
   &=&  -\frac{1}{16\pi G}(\sigma+\frac{\alpha H}{\mu}+\frac{F}{\mathrm{m}^2})\int_{0}^{2\pi}\bigg\{T(\phi)\mathcal{M}(\phi)+2Y(\phi)\mathcal{L}(\phi)\bigg\}d\phi\nonumber \\
   &+& \frac{1}{16\pi G \mu}\int_{0}^{2\pi}Y(\phi)\mathcal{M}(\phi)d\phi
\end{eqnarray}
where $Y(\phi)$, and $T(\phi)$ are given by (3.58), (3.61) respectively.
Using the holographic disctionary (\ref{var.rel}), we find that the gravitational charge (\ref{F-charge GMMG2}) evaluated at the asymptotic boundary agrees with the modular Hamiltonian (\ref{H in FT-therm}) in the field theory side. This is another evidence for the $flat_3/BMSFT$ correspondence. As shown in \cite{Apolo-2}, the gravitational charge along the swing surface equals the entanglement entropy. Thus, we can generalize the first law of entanglement entropy (\ref{match1}) to the thermal BMSFTs.

\section{Conclusion}\label{sec:4}
In this paper we studied a new class of non-AdS holography. This is the $Flat_3/BMSFT$ holography in the framework of GMMG model \cite{Setare-Adami Nucl 2018}. By applying the Bondi fall-off conditions \cite{Bondi-1,Sach-1,Sach-2}, we found the Killing vectors (\ref{Killing GMMG}) of the isometries that preserve the asymptotic form of the metric in (\ref{B.C.}). The conserved charges corresponding to these vectors are dual to the modular Hamiltonian of the field theory at the boundary.

To find the modular Hamiltonian \cite{Haag-bk,Casini-1}, we considered a BMS-invariant field theory at the boundary. The BMSFTs are considered non-thermal on the plane and thermal on the cylinder. To compute the vacuum modular Hamiltonian, the generalized Rindler transformation was used. In this method, the vacuum state was defined on an interval $\mathcal{I}$ that was mapped to a Rindler spacetime under a symmetry transformation. The modular flow generator (\ref{zeta-1}) was considered as a linear combination of the vacuum symmetry generators. Using the conditions on the Rindler transformations, the coefficients (\ref{aj2}) and (\ref{bj2}) of the modular flow generator were computed.

To compute the modular Hamiltonian,  the vacuum symmetry generators $l_j$ and $m_j$ were replaced with the corresponding conserved charges $\mathcal{L}_j$ and $\mathcal{M}_j$. By this replacement the modular flow generator (\ref{zeta-1}) was replaced with the modular Hamiltonian (\ref{H in FT}). Using the equal forms of the charge algebras (\ref{BMS algebra}) of the two sides of the holography, we found the dictionary (\ref{var.rel}) between the variables of the bulk and the boundary. Using this dictionary of the holography, we showed that the gravitational charge (\ref{F-charge GMMG}) is the same as the
modular Hamiltonian (\ref{H in FT}). Finally we were able to show the first law of entanglement in (\ref{match1}).

We first did these steps for a Poincare patch (\ref{Poin}) of GMMG model and a zero temperature BMSFT on a plane (\ref{interval}) on two sides of the holography. Then, we considered a general case in which the metric has a flat space cosmological form (\ref{FSC}). In this case, the holographic dual of the gravity is a finite temperature BMSFT on a cylinder (\ref{therm. cyl}). Using the asymptotic form of the Killing vectors of FSC (\ref{Kil-FSC-2}), the modular Hamiltonian for a single interval on a thermal BMSFT was computed in (\ref{H in FT-therm}). The computations associated with the conserved charges of GMMG model were repeated in (\ref{F-charge GMMG2}). In this general case, the first law of holographic entropy was obtained too.
It is interesting that our results for GMMG model reduces to the result for Einstein gravity in $2+1$ dimension, when $\mu , m^2 \to \infty $, and $\sigma=-1$. For example one can explicitly see that our expression for variation of conserved gravitational charge in Eq(3.17) reduce to the
 expression (3.51)  in the paper \cite{Apolo} for Einstein gravity in mentioned limiting cases. These limiting cases, are a place in parameter space of model, where GMMG model reduces to the Einstein gravity.

\end{document}